\documentclass[aps,notitlepage,nofootinbib,11pt,tightenlines]{revtex4}
\usepackage{graphicx}
\usepackage{float}
\usepackage[T1]{fontenc}
\usepackage[latin1]{inputenc}
\usepackage{amssymb}
\usepackage{ae}
\usepackage{aecompl}

\include{epsf}

\newcommand{\bea}{\begin{eqnarray}}
\newcommand{\eea}{\end{eqnarray}}

\begin{document}


\title{Perturbative and Nonperturbative Kolmogorov Turbulence in a Gluon Plasma}

\author{M.E. Carrington}
\email{carrington@brandonu.ca}
\affiliation{Department of Physics, Brandon University, Brandon, Manitoba, R7A 6A9 Canada\\ and \\  Winnipeg Institute for Theoretical Physics, Winnipeg, Manitoba }
\author{A. Rebhan}
\email{rebhana@hep.itp.tuwien.ac.at}
\affiliation{Institute for Theoretical Physics, Wiedner Hauptstra\ss e 8-10, Vienna University of Technology, A-1040 Vienna}

\begin{abstract}
In numerical simulations of nonabelian plasma instabilities in the hard-loop approximation, a turbulent spectrum has been observed that is characterized by a phase-space density of particles $n(p)\sim p^{-\nu}$ with exponent $\nu\simeq 2$, which is larger than expected from relativistic $2\leftrightarrow 2$ scatterings. Using the approach of Zakharov, L'vov and Falkovich, we analyse possible Kolmogorov coefficients for relativistic $(m\!\ge\!4)$-particle processes, which give at most $\nu=5/3$ perturbatively for an energy cascade. 
We discuss nonperturbative scenarios which lead to larger values. As an extreme limit we find the result $\nu=5$  generically in an inherently nonperturbative effective field theory situation, which coincides with results obtained by Berges et al.\ in large-$N$ scalar field theory. 
If we instead assume that scaling behavior is determined by Schwinger-Dyson resummations such that the different scaling of bare and dressed vertices matters, we find that intermediate values are possible. We present one simple scenario which would single out $\nu=2$.
\end{abstract}
\maketitle

\section{Introduction}
Turbulence in nonequilibrium field theory \cite{zak} 
has been studied intensively
in scalar field theories \cite{micha,jbLET,BergesHoff,Berges:2010ez} and more recently in the
context of quantum chromodynamics (QCD)
\cite{Mueller:2006up,arnoldmoore2,Berges:2008mr}. 
The nonequilibrium dynamics of QCD is of particular interest for the early
evolution and thermalization of quark-gluon plasma \cite{botup}
in a heavy ion collision.
At least in a weakly coupled quark-gluon plasma, anisotropic distributions
of particles give rise to plasma instabilities \cite{mrow} and exponential
growth of collective fields \cite{romstr,lenag1,Rebhan:2009ku}. These instabilities provide an infra-red source of energy
which can cascade
to higher energies by nonabelian interactions once the corresponding
field modes have grown to nonperturbatively large amplitudes.
For a stationary anisotropic plasma,
as considered in the hard-loop framework, a quasi-stationary state with scale
independent transport to the ultra-violet can form. This state is characterized by
a linear growth of energy \cite{Arnold:2005vb,Rebhan:2005re,Bodeker:2007fw}, with the formation of a power-law spectrum \cite{arnoldmoore2,IRS}
indicative of the appearance of Kolmogorov turbulence \cite{zak,micha}.

In this paper, we employ the approach of Zakharov, L'vov and Falkovich \cite{zak,micha} to find possible Kolmogorov exponents in a relativistic gauge theory with an approximately isotropic power law spectrum:
\bea
\label{nuDefn}
1 \ll n(p)\sim p^{-\nu}~~{\rm for} ~~p^*<p<\Lambda(t)\,,
\eea 
which for $\nu>1$
represents a cascade from soft perturbatively unstable modes at scales below $p^*$ towards the ultra-violet.
Distributions of this form with $\nu\simeq 2$ have
been observed in hard-loop real-time lattice simulations of
an anisotropic purely gluonic plasma \cite{arnoldmoore2,IRS}.
Using
classical-statistical lattice simulations which do not allow for a separation of scales of hard particles
and soft collective modes,
Ref.~\cite{Berges:2008mr} has found a significantly lower
value at large times, $\nu\approx 4/3$, indicative of perturbative
Kolmogorov scaling. However, these latter simulations may not yet
have probed the parametrically separated
infrared regime responsible for the different behavior
seen in the hard-loop simulations.

Taking into account the renormalizable classically scale-invariant
interactions of QCD, we obtain the results $\nu=5/3$ and $\nu=4/3$
for energy and particle cascades, respectively, which are familiar from
scalar $\phi^4$ theory for $2\leftrightarrow 2$ processes.
However it is well known that in the infrared regime resummations can play an important role. This is especially true in gauge theories, where all-order
resummations lead to effectively non-local interactions \cite{PA,AMY}. In addition to
perturbative scaling exponents, there may also exist nonperturbative ones which dominate at lower momentum scales. We therefore consider the effect of 
nonperturbative propagator and 
vertex resummations that correspond to a modified
scaling law for dressed quantities.

If we determine this modified law
by assuming the same scaling at all loop orders, we obtain a Kolmogorov exponent $\nu=5$ for an energy cascade generically in an inherently nonperturbative effective field theory situation.
This coincides with the nonperturbative infrared fixed point found previously
by Berges et al.\ in large-$N$ scalar field theory \cite{jbLET,Berges:2010ez},
but not observed in any numerical simulation of nonabelian gauge theory
so far. Assuming 
instead that the scaling behavior of full vertex functions is determined by Schwinger-Dyson resummations such that the different scaling of bare and dressed vertices matters, we find a range of possible intermediate values, and we present a scenario which yields $\nu=2$.

\section{Zakharov Transformations}
\label{zakSection}

We can formulate the condition for stationary, scale-independent
transport in terms of detailed balance, by requiring that 
`gain' and `loss' terms balance each other, which means that 
the rate at which energy goes into the cascade is equal to the rate at which it flows out. 
In Appendix \ref{michaSection} we show that this condition is equivalent to the requirement that flux be scale invariant (which is the approach used in Ref. \cite{micha}).  
The condition for detailed balance can be written using the closed time path (CTP) formalism (see e.g.\
Ref.~\cite{stanM}). 
We make the following definitions:
\bea
\label{ret-xspace}
&& G^{ret}(x,y) =
-i\Theta(x_0-y_0) \langle [\phi(x),\phi(y)] \rangle =G_{11}(x,y) - G_{12}(x,y)\;, \\[2mm] 
\label{adv-xspace}
&& G^{adv}(x,y) = 
i\Theta(y_0-x_0) \langle [\phi(x),\phi(y)] \rangle = G_{11}(x,y) - G_{21}(x,y)\;,
\nonumber\\[2mm] 
&& \label{sym-xspace}
G^{sym}(x,y) = 
-i \langle \{ \phi(x),\phi(y) \}\rangle = G_{11}(x,y) + G_{22}(x,y)
 \;,\nonumber\\[2mm]
 && \rho(x,y)  = \langle [\phi (x), \phi (y)] \rangle 
 = i\big(G^{ret}(x,y)  -  G^{adv}(x,y)\big) =id(x,y)\;.\nonumber
\eea
We use analogous expressions for the self energies:
\bea
\Pi^{ret}(x,y)=\Pi_{11}(x,y) + \Pi_{12}(x,y)\;,~~\Pi^{adv}(x,y)=\Pi_{11}(x,y) + \Pi_{21}(x,y)\;,\\[2mm]
\Pi^{d}(x,y)=\Pi^{ret}(x,y)-\Pi^{adv}(x,y)\;,~~
\Pi^{sym}(x,y)=\Pi_{11}(x,y) + \Pi_{22}(x,y)\;.\nonumber
\eea
Using these definitions, the condition of stationary transport can be written:
\newpage
\bea \label{Zdefn}
Z&=& \theta(p_0)\int d^3p\,~ \rho(P)\,\cdot\, \bigg(p \,\cdot\,\frac{n(p)}{dt}\bigg)= 0 \\
\rightarrow && \theta(p_0) \!\int\! d^3p\, \big(\Pi^{sym}(P) d(P)-  \Pi^d(P) G^{sym}(P)\big)=0\,.\nonumber
\eea
Throughout this paper we use a shorthand notation a function of $p_0$ and $p$: $f(P):=f(p_0,p)$.

All momenta in the cascade are considered soft relative to the ultra-violet end of the cascade. 
Consequently, we can try to solve (\ref{Zdefn}) by looking for scaling solutions that correspond to infra-red fixed points.
In a perturbative regime of QCD,
the individual building blocks of a given amplitude scale according to:
\bea
\label{scaleDefn}
&& G(\lambda P) = |\lambda|^{-2}\,G(P)~~\text{for $G$ =${\rm Re}\,[G^{ret}]$ or ${\rm Re}\,[G^{adv}]$}\,, \\
&& d(\lambda P) = {\rm sgn}\,(\lambda)\;|\lambda|^{-2}\,G(P)\,, \nonumber\\
&& G^{sym}(\lambda P) = |\lambda|^{-2-\nu}\,G^{sym}(P)\,,~~n(\lambda P) = |\lambda|^{-\nu}n(P)\,, \nonumber\\
&& U(\lambda P) = {\rm sgn}\,(\lambda)\;|\lambda|\, U(P)\,,~~V(\lambda P) = V(P)\,,\nonumber
\eea
where $U$ represents the 3-point vertex and $V$ the 4-point vertex. 

\subsection{1 $\leftrightarrow$ 2 processes}\label{12processes}
\label{2to1}

We begin
by using in  (\ref{Zdefn}) a 1-loop self energy diagram  which corresponds to
a 1 $\leftrightarrow$ 2 process, even though this process is kinematically forbidden 
for massless particles. We use the method in Ref. \cite{TF} to perform the sums over the CTP indices and obtain:
\begin{widetext}
\bea
\label{1loop}
&& Z\sim \theta(p_0)\int d^3p \int dQ\,dL\; \delta(P^2)\,\delta(Q^2)\,\delta(L^2)\delta^4(P+Q+L)U(Q,P,L)\;U(-L,-P,-Q) F(p_0,q_0,l_0)\,,\nonumber\\
&& ~~~ F(p_0,q_0,l_0) = n(p_0)n(q_0)n(l_0)\left[\frac{1}{n(p_0)}+\frac{{\rm sgn}\,(q_0)}{n(q_0)}+\frac{{\rm sgn}\,(l_0)}{n(l_0)}\right]\,.
\eea
\end{widetext}
The  integral in (\ref{1loop}) is identically zero, because the product of delta functions has no support. We proceed with the calculation to establish our notation.

Following Ref. \cite{zak} we perform two different transformations and add the results to the original expression.
The substitutions are ($\lambda_2=p_0/q_0^\prime$, $\lambda_3=p_0/l_0^\prime$):
\bea
\label{shift1loop1}
\{q_0\rightarrow  \lambda_2 p_0,l_0\rightarrow  \lambda_2 l_0^\prime,{\bf p}\rightarrow \lambda_2 {\bf q}^\prime,{\bf q}\rightarrow \lambda_2 {\bf p}^\prime,{\bf l}\rightarrow \lambda_2 {\bf l}^\prime\}\,,\\
\label{shift2loop1}
\{q_0\rightarrow  \lambda_3 q_0^\prime,l_0\rightarrow  \lambda_3 p_0,{\bf p}\rightarrow \lambda_3 {\bf l}^\prime,{\bf q}\rightarrow \lambda_3 {\bf q}^\prime,{\bf l}\rightarrow \lambda_3 {\bf p}^\prime\}\,,\nonumber
\eea
and they correspond respectively to transforming the set of variables:
\bea
\{P,Q,L\} \rightarrow \lambda_2 \{Q^\prime,P^\prime,L^\prime\}\,,~~
\{P,Q,L\} \rightarrow \lambda_3 \{L^\prime,Q^\prime,P^\prime\}\,.
\eea
where we have defined $P^\prime=(p_0,{\bf p}^\prime)$.
The integrand in (\ref{1loop}) is symmetric under the interchange of any two arguments, so we can combine the results from both substitutions with the original expression to obtain a result of the form:
\bea
\label{1loopGEN}
&&\!\!\!\!\!Z \sim \theta(p_0) \int d^3p^\prime \int dQ^\prime \,dL^\prime(1+{\rm sgn}\,\big(\lambda_2)\;|\lambda_2|^\Delta + +{\rm sgn}\,(\lambda_3\big)\;|\lambda_3|^\Delta){\cal F}(P^\prime,Q^\prime,L^\prime)\,.
\eea
The exponent $\Delta=4-2\nu$ and is obtained from:
\bea
\lambda^{3\cdot 3+1+2-4}|\lambda|^{-6+2-2\nu} = \lambda^8|\lambda|^{-4-2\nu}=|\lambda|^{4-2\nu}=|\lambda|^\Delta\,.
\eea
The exponent of the first term is: ${3 \cdot 3}$  from the integrals over the spatial momenta $\{{\bf p},{\bf q},{\bf l}\}$,  1 from the $l_0$ integral, 2 from the $q_0$ integral (since $q_0 \to \lambda_2p_0 = p_0^2/q_0^\prime)$ and $-4$ from the overall momentum conserving delta function. The exponent of the second term is: $-6$ from the on-shell delta functions, 2 from the vertex functions, and $-2\nu$ from the distribution functions. 
For $\Delta=-1$ the bracket in the first line of  (\ref{1loopGEN}) becomes:
\bea
1+\frac{1}{\lambda_2}+\frac{1}{\lambda_3} = \frac{1}{p_0}\bigg[p_0+q^\prime_0+l^\prime_0\bigg] = 0\,,
\eea
where we have used  that the integrand in (\ref{1loop}) contains a delta function of the form $\delta(p_0+q_0^\prime+l_0^\prime)$.  

If $Z$ were not identically zero anyway, due to kinematic constraints,  $\Delta=-1$ would have given a scaling solution of $\nu=5/2$. 
The function $F(p_0,q_0,l_0)$ in (\ref{1loop}) is zero for $\nu=1$, and therefore $\nu=1$ is also a solution of $Z=0$.  This is not a scaling solution however, because the quantity in square brackets in (\ref{1loopGEN}) is not zero. The solution $\nu=1$ corresponds to the  equilibrium thermal distribution function, and the equality $F(p_0,q_0,l_0,r_0)\big|_{\nu=1}=0$  is just the KMS condition.

\subsection{2 $\leftrightarrow$ 2 processes}
\label{2to2}

Since the contribution to $Z$ from the 1-loop diagram is identically zero due to kinematic constraints, we turn to the 2-loop self energy.
If we include all 2-loop self energy diagrams in  (\ref{Zdefn}), corresponding
to all possible 2 $\leftrightarrow $ 2 processes, we obtain:
\begin{widetext}
\bea
\label{2loop}
&& Z\sim \theta(p_0)\int d^3p \int dQ\, dR\,dL\; \delta(P^2)\,\delta(Q^2)\,\delta(L^2)\delta(R^2)\delta^4(P+Q-L-R)\big|\mathcal M\big|^2 F(p_0,q_0,l_0,r_0)\,,\nonumber\\
&& ~~~F(p_0,q_0,l_0,r_0) = n(p_0)n(q_0)n(l_0)n(r_0)\left[\frac{1}{n(p_0)}+\frac{{\rm sgn}\,(q_0)}{n(q_0)}-\frac{{\rm sgn}\,(l_0)}{n(l_0)}-\frac{{\rm sgn}\,(r_0)}{n(r_0)}\right]\,,
\eea
\end{widetext}
where the square of the matrix element $\big|\mathcal M\big|^2$ is a symmetric function of the Mandelstam variables:
$s=(P+Q)^2,$ $t=(P-L)^2$ and  $u=(P-R)^2$. 
In order to find the scaling solutions to (\ref{Zdefn})  
we perform three different substitutions and add the results to the original expression: 
These three shifts are:
\bea
\label{shift2}
&& \{ q_0 \rightarrow \lambda_2 p_0, ~l_0 \rightarrow \lambda_2 l^\prime_0, ~r_0 \rightarrow \lambda_2 r^\prime_0, ~{\bf p} \rightarrow \lambda_2{\bf q}^\prime, ~{\bf q} \rightarrow \lambda_2{\bf p}^\prime, ~{\bf l} \rightarrow \lambda_2{\bf l}^\prime, ~{\bf r} \rightarrow \lambda_2{\bf r}^\prime\}\,;~~\lambda_2=\frac{p_0}{q_0^\prime}\,,\\
&& \{ q_0 \rightarrow -\lambda_3 q^\prime_0, ~l_0 \rightarrow \lambda_3 p_0, ~r_0 \rightarrow -\lambda_3 r^\prime_0, ~{\bf p} \rightarrow \lambda_3{\bf l}^\prime, ~{\bf q} \rightarrow -\lambda_3 {\bf q}^\prime, ~{\bf l} \rightarrow \lambda_3 {\bf p}^\prime, ~{\bf r} \rightarrow -\lambda_3 {\bf r}^\prime\}\,;~~\lambda_3=\frac{p_0}{l_0^\prime}\,,\nonumber \\
&& \{ q_0 \rightarrow -\lambda_4 q^\prime_0, ~l_0 \rightarrow \lambda_4 p_0, ~r_0 \rightarrow -\lambda_4 l^\prime_0, ~{\bf p} \rightarrow \lambda_3{\bf r}^\prime, ~{\bf q} \rightarrow -\lambda_4 {\bf q}^\prime, ~{\bf l} \rightarrow \lambda_4 {\bf p}^\prime, ~{\bf r} \rightarrow -\lambda_4 {\bf l}^\prime\}\,;~~\lambda_4=\frac{p_0}{r_0^\prime}\,.\nonumber
\eea
Using the notation $P'=(p_0,\mathbf p')$ we can rewrite  (\ref{shift2}) in the form:
\bea
&& \{P,Q,L,R\} \rightarrow \lambda_2 \{Q^\prime,P^\prime,L^\prime,R^\prime\}\,,\\[2mm]
&& \{P,Q,L,R\} \rightarrow \lambda_3 \{L^\prime,-Q^\prime,P^\prime,-R^\prime\}\,,\nonumber\\[2mm]
&& \{P,Q,L,R\} \rightarrow \lambda_4 \{R^\prime,-Q^\prime,P^\prime,-L^\prime\}\,.\nonumber
\eea
Combining the contributions from all three shifts with the original expression, the integral in (\ref{2loop}) becomes\footnote{We use $\Delta$ for the scaling exponent of $Z$ throughout this paper, without introducing subscripts to distinguish results that  correspond to different choices for the self-energies in  (\ref{Zdefn}).}:
\bea
\label{Z2loop}
 Z \sim \theta(p_0)\int d^3p^\prime \int dQ^\prime \,dL^\prime \,dR^\prime\;{\cal F}(P^\prime,Q^\prime,L^\prime,R^\prime)\big[1 +{\rm sgn}\,(\lambda_2)\;|\lambda_2|^\Delta  - {\rm sgn}\,(\lambda_3)\;|\lambda_3|^\Delta  - {\rm sgn}\,(\lambda_4)\;|\lambda_4|^\Delta
\big]\,,\nonumber\\
\eea
with $\Delta =4-3\nu$ obtained from:
\bea
\lambda^{12}|\lambda|^{-8-3\nu}= |\lambda|^{4-3\nu} = |\lambda|^\Delta\,.
\eea
The first exponent is  $(4\cdot 3+4)-4$ from the Jacobian of the transformation and the 4-dimensional delta function, and the second exponent receives contributions: $-8$ from the on-shell delta functions  and $-3\nu$ from the distribution functions. 

Using the definitions of the $\lambda$ variables in  (\ref{shift2}) 
together with momentum conservation $P+Q=L+R$,
a scaling solution to  (\ref{Zdefn}) is obtained:
\bea
-1= \Delta = 4-3\nu ~~\Rightarrow \nu=\frac{5}{3}\,.
\eea
This result is the same in scalar theories when 2 $\leftrightarrow$ 2 processes are taken to give the dominant contribution to the collision integral. 
This indicates a universality corresponding to an insensitivity to the particular theory under consideration. 

As in section \ref{2to1}, we have an algebraic solution when $\nu=1$.  This solution corresponds to the equilibrium KMS condition, and is not a scaling solution, because the quantity in square brackets in (\ref{Z2loop}) is not zero.
Note also that there is a scaling solution for $\Delta=0$ ($\nu = 4/3$) if we make use the restriction imposed by the delta functions in (\ref{2loop}) to consider only 2 $\rightarrow$ 2 processes, for which two of $\lambda_2$, $\lambda_3$ and $\lambda_4$ are positive. This solution corresponds physically to the scale invariance of particle flux (see Appendix \ref{michaSection}). 
For a relativistic field theory, where particle number is not conserved, the physically relevant solution would seem to be the one that corresponds to the scale invariance of energy flux. Remarkably however, the numerical results in classical statistical field theory simulations of a nonabelian gauge theory in Ref.\ \cite{Berges:2008mr} appear to favor $\nu=4/3$. On the other hand, simulations of nonabelian plasma instabilities within the hard-loop framework have found $\nu\simeq 2$ in a momentum regime that is parametrically separated from the hard scale of plasma constituents \cite{arnoldmoore2,IRS} and perhaps out of reach of the simulations of Ref.\ \cite{Berges:2008mr}.


\subsection{Zakharov Scaling at Arbitrary Loop Order}
\label{lloopSection}

We can generalize the results of the previous two sections and obtain an expression for the scaling exponent of an $l$-loop amplitude. 
In the next section we show how to use this result to find solutions to  (\ref{Zdefn}) that  correspond to infinite resummations of vertex corrections. 
In the CTP formalism, $n$-point functions have $2^n-1$ independent components, and integral equations for $n$-point functions couple different components. However, the situation becomes vastly simplified in the statistical limit where we only need to keep terms that carry the greatest number of distribution functions, which is equal to the number of loops. 
We define: $l$  = number of loops, $I$ = number of internal lines,
$v_3$  =  number of cubic vertices, $v_4$=number of quartic vertices, and $E$  = number of external legs.
We denote the complete set of momentum arguments for the external legs: $P_i:=\{P_1,P_2,\dots P_E\}$, and the momentum arguments for the loop variables: $L_j:=\{L_1,L_2\dots L_l\}$. An arbitrary amplitude can be written symbolically:
\bea
{\cal A}(P_i) = \bigg[\int dL_j\bigg]^{l}\;\big(G^{sym}(P_i,L_j)\big)^{l} \; \big(G(P_i,L_j)\big)^{I-l}\;\big(U(P_i,L_j)\big)^{v_3}\;\big(V(P_i,L_j)\big)^{v_4}.\nonumber
\eea
Defining $L_j=\lambda L_j^\prime$ and using (\ref{scaleDefn}) we obtain:
\bea
\label{Adefn}
{\cal A}(\lambda P_i) &&= \big({\rm sgn}\,(\lambda)\big)^{E}\,|\lambda|^{-\gamma}{\cal A}(P_i)\,,~~\gamma = 2I+\nu\cdot l-v_3-4l\,.
\eea
The exponent on the sign function is obtained from the fact that an amplitude with $E$ odd will always correspond to $v_3$ odd. 
Using the topological constraints:
\bea
\label{topo}
&&l=I-(v_3+v_4)+1\,,~~E=4v_4+3v_3-2I\,,
\eea
the exponent $\gamma$ can be rewritten in different forms:
\bea
\label{gammaBare}
\gamma = \nu - 4 + E
+ \frac{v_3}{2}\nu
+ v_4\nu = E-4+l\nu\,.
\eea

We can find the Zakharov scaling solution of (\ref{Zdefn}) with $l$-loop self-energies by following the method in the previous section, provided we choose a set of diagrams that have the appropriate symmetry under 
permutations
of the momentum arguments of the on-shell propagators. 
Note that in the  2-loop calculation discussed in section \ref{2to2}, the complete set of 2-loop diagrams produces the fully symmetric matrix element in Eq. (\ref{2loop}). At higher loop orders we expect that the set of diagrams with the correct symmetry is equal to the full set of diagrams at that order. 
The exponent $\Delta$ is obtained from $-\gamma$ by including the two factors (see Eq. (\ref{Zdefn})):

\noindent (1) $\lambda^{3+1}=|\lambda|^4$ where the 3 is from the integral over 3-momenta, and the 1 is from the  integral over the zeroth momentum component that is interchanged with the external momentum by the Zakharov transformation.

\noindent (2) $|\lambda|^{-(2+\nu)}$ from the extra factor proportional to $G^{sym}(p)$.
 
\noindent Combining these factors the solution for the scaling exponent is obtained from:
\bea
\label{zakBare}
&& -1=\Delta  = 4-(2+\nu) -\gamma \,.
\eea
Substituting (\ref{gammaBare}) into (\ref{zakBare}) and using $E=2$ we obtain:
\bea
\label{nu1}
~~\Rightarrow ~~&&   \nu = \frac{5}{1+l} = \frac{5}{1+\frac{1}{2}(v_3+2v_4)}\,.\nonumber
\eea
Note that $l=2$ gives $\nu=5/3$ as discussed above. 
Higher loop orders, which correspond to processes involving $m=l+2>4$ particles, are seen to be
subdominant. 

In section \ref{2to2} we saw that there is also a scaling solution for $\Delta=0$ at the 2-loop level corresponding to a particle cascade. 
With arbitrary loop orders contributing, we do not expect $\Delta=0$ to be a solution. This would require that only $n \leftrightarrow n$ processes contribute, which seems unlikely in a relativistic field theory, where particle number is not conserved.


\section{Resummations}

\subsection{Resummed Vertices}
\label{vertexSection}

In section \ref{zakSection} we have assumed bare 3- and 4-vertices and standard
scaling behavior of propagators, except for modified statistical factors.
When calculating transport coefficients and production rates in gauge theories, the non-locality of effective interactions requires us to include diagrams that correspond to infinite series of interactions with soft particles \cite{PA,AMY}. Traditionally, these diagrams are packaged into resummed vertex corrections. 
We expect that the Zakharov scaling exponents will also receive contributions from processes involving an unlimited number of particles.
We now consider the possibility that these processes can be included using resummed propagators and vertices.

First we remind the reader of the structure of a traditional ladder resummation. 
The basic idea is that although each term in the ladder series is formally of higher order in the coupling constant, kinematic enhancements change the naive power counting. The dominant contribution to each ladder diagram comes from the region of phase space that corresponds to hard rails (to maximize phase space) and the soft rungs (to obtain a Bose enhancement). The end result is that all diagrams in the series contribute at the same order, which means that all ladder diagrams have to be resummed to obtain the leading order contribution. 

The structure of the calculation of a scaling exponent is quite different. 
One basic difference is that all lines carry momenta that correspond to soft particles in the cascade, and thus it is not possible to separate two scales which could play the role of hard rails and soft rungs. 
We note that the method that we have described to find scaling solutions requires that all distribution functions be approximated by their classical forms, so that the integrand satisfies a scaling law.  This means that, in any scaling calculation, it is not possible to treat some momenta as parametrically harder than others, without losing the symmetry that is needed to do the calculation. 

Another difference is discovered by considering the scaling behaviour of an $n$-loop amplitude calculated with bare lines and vertices. From equation (\ref{Adefn}) we see that in the infra-red where $|\lambda|$ is large, we have ${\cal A}(P_i)$  large for $\gamma$ positive and small for $\gamma$ negative.
From  (\ref{gammaBare}) we have:
\bea
\label{gammaBare2}
\gamma^{\rm tree} = E-4\,,~~
\gamma^{\rm 1-loop}  = E-4+\nu \,,
\eea
and therefore for $\nu$ positive, a 1-loop vertex calculated with bare lines and vertices is larger than the corresponding tree vertex. 
 This situation is completely different from that encountered in ladder resummations,  where the resummation involves an infinite series of diagrams which all contribute at the same order. In our case, higher loop graphs are larger than lower loop ones, and any traditional resummation will fail to converge. This is similar to the well known Linde problem, and is therefore expected at very soft momentum scales. 

The key point is that in a scaling calculation, we do not need to actually perform the resummation, but only to extract effective scaling exponents that correspond to resummed lines and vertices. 
We define a
resummed propagator $\tilde G(P)$, 3-vertex $\tilde U(P_i)$ and 4-vertex $\tilde V(P_i)$ with modified scaling laws (compare with  equation (\ref{scaleDefn})):
\bea
\label{gammaEx}
&& \tilde G(\lambda P) = |\lambda|^{-2-x_2\nu}\,\tilde G(P)~~\text{for $\tilde G$ =${\rm Re}\,[\tilde G^{ret}]$ or ${\rm Re}\,[\tilde G^{adv}]$}\,, \\
&& \tilde d(\lambda P) = {\rm sgn}\,(\lambda)\;|\lambda|^{-2-x_2\nu}\,\tilde G(P)\,, \nonumber\\
&& \tilde G^{sym}(\lambda P) = |\lambda|^{-2-(1+x_2)\nu}\,\tilde G^{sym}(P)\,, \nonumber\\
&& \tilde U(\lambda P) = {\rm sgn}\,(\lambda)\;|\lambda|^{1+x_3\nu}\, \tilde U(P)\,,~~\tilde V(\lambda P) = |\lambda|^{x_4\nu}\,\tilde V(P)\,.\nonumber
\eea
In addition, we include effective $(n>4)$-vertices with scaling laws:
\bea
\tilde V_n(\lambda P_i)= \big({\rm Sign}\,(\lambda)\big)^n \, |\lambda|^{-(n-4-x_n\nu)}\tilde V_n(P_i)\,.
\eea
The
topological relations obtained by generalizing (\ref{topo}) are:
\bea
\label{topoBG}
l=I+1-\sum_{k=3}^\infty v_k\,,\quad
E=\sum_{k=3}^\infty k\, v_k-2I\,.
\eea
Note that setting $x_n=0$ we recover the scaling laws of the bare quantities (Eq. (\ref{scaleDefn})), apart from the fact that there are no bare $n$-point-vertices
with $n>4$.

Equation (\ref{Adefn}) gives the scaling exponent for an arbitrary amplitude calculated with bare lines and vertices. The generalization of this result for amplitudes with resummed lines and vertices is:
\bea
\label{gamma0}
\gamma_x =  I(2+x_2\nu)+(\nu-4) l-v_3(1+x_3 \nu)-v_4(0+ x_4 \nu) -\ldots\,,\nonumber\\
\eea
and using (\ref{topoBG}) to eliminate $I$ and $l$ 
we find:
\bea
\label{gamma2}
\gamma_x = \nu - 4 + \frac{E}{2} (2-\nu-x_2\nu)
+ \frac{v_3}{2}(1+3 x_2-2 x_3 )\nu
+ v_4(1+2x_2-x_4)\nu+\ldots
\eea

\begin{figure}[H]
\begin{center}
\includegraphics[width=10cm]{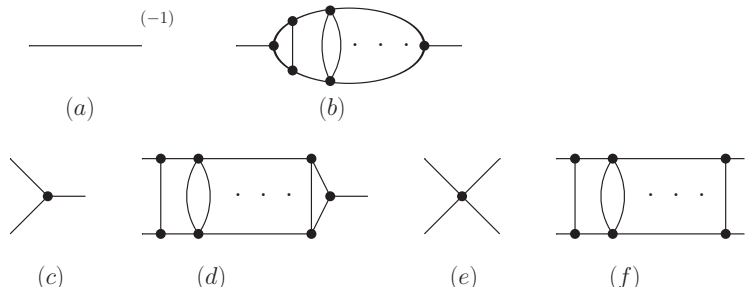}
\end{center}
\caption{\label{selfconsisALL}Graphs that are required to have the same scaling exponent. The dots indicate any number of additional rungs.}
\end{figure}

If an all-order nonperturbative resummation is indeed able to change
the scaling behavior of an arbitrary amplitude, we expect that the scaling
exponent of the inverse propagator (a) in Fig. \ref{selfconsisALL} equals that of graph (b), and that the 3-point graph (c) has the same scaling exponent as graph (d),
and that the 4-point graph (e) has the same scaling exponent as graph (f).
\par
\noindent This leads us to the conjecture that a nonperturbative fixed point
is characterized by values of $\{x_2,x_3,x_4,\ldots\}$ for which
$\gamma$ only depends on the number of external legs, and not
on $v_3$ and $v_4$ and higher. 
We comment that this 
includes not only ladder graphs, as shown in Fig. \ref{selfconsisALL}, but also graphs which correspond to crossed ladders, and even graphs with non-ladder structure. As discussed previously, the fact that our definition of resummed lines and vertices includes non-ladder topologies is consistent with the physics of the cascade, in which all particles are soft. 
Imposing the condition that  (\ref{gamma2}) is independent of the number of vertices fixes the scaling of effective vertices in terms of $x_2$ according to:
\bea\label{x342}
x_3=\frac12(1+3 x_2)\,,\quad x_4=1+2x_2\,,\quad\ldots
\eea

We could attempt to find the scaling exponent directly from the generalization of (\ref{zakBare}):
\bea
\label{zakBareBG}
0=3+1-\Delta-\big((2+x_2\nu)+\nu\big)-\gamma_x\bigg|_{E=2}\,.
\eea
Substituting  (\ref{gamma2}) and (\ref{x342}) directly into  (\ref{zakBareBG}) we find that the dependence on $x_2$ cancels exactly, and we obtain $\nu=5$, at any loop order. The cancellation of $x_2$ can be seen directly using simple power counting, and the fact that the integrand in the second line of (\ref{Zdefn}) has the structure $n(p)G(p)\Pi(p) \sim n(p)G(p)(G^{-1}(p)-G_0^{-1}(p))$.
We expect that the result is independent of the number of loops, since the resummation procedure we have used is just the condition that diagrams with an arbitrary number of loops contribute at the same order. 

The result (in 3+1 dimensions) is:
\bea
\label{fiveRes}
\nu=
5 & \text{for $\Delta=-1$  (energy cascade)} \,.
\eea
This result coincides with the nonperturbative infrared fixed points found in Refs. \cite{jbLET,Berges:2010ez} in (large-$N$) scalar field theory. Note also that result in (\ref{fiveRes}) contains a prediction of the values of the $x_i$'s which could be checked numerically. 
As discussed above, the actual scaling of propagators and vertices does not play a role, which suggests that this extremely large result is a universal limit. The result depends on the implicit assumption of the existence of an effective theory in the infrared where exclusively dressed vertex functions govern the dynamics.

\begin{figure}[H]
\begin{center}
\includegraphics[width=13cm]{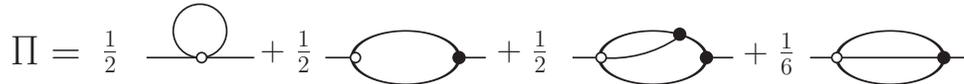}
\end{center}
\caption{\label{SD-SE}The Schwinger-Dyson equation for the gluon self-energy. Dressed vertices are indicated by filled vertex dots and propagator lines are assumed to be dressed. 
}
\end{figure}

In situations where the specific microscopic dynamics is relevant, the diagrammatics is instead governed by Schwinger-Dyson equations which  combine bare and dressed vertices. If we assume that $\gamma_x|_{E=2}$ is characterised by the diagrams shown in Fig.~\ref{SD-SE}, we should make the substitution
\bea
v_k \to n_k+x_k\nu n_k'
\eea
in (\ref{gamma0}), where $n_k$ and $n_k'$ are the numbers of bare and dressed $k$-vertices, respectively. 

\par

The second diagram in Fig.~\ref{SD-SE} corresponds to $n_3=n_3'=1$ and yields:
\bea
\nu=\frac{5}{2+3x_2-x_3} \quad\mbox{for $\Delta=-1$}\,.
\eea
Using the scaling relations (\ref{x342}), which express nonperturbative behaviour in the sense that all loop orders contribute equally to dressed quantities, we obtain:
\bea
\label{10over3}
\nu=\frac{10}{3(1+x_2)} \quad\mbox{for $\Delta=-1$}.
\eea
The tadpole diagram in the Schwinger-Dyson equation does not have an imaginary part. The third diagram in Fig.~\ref{SD-SE} gives $\nu=5/(3+5x_2-2x_3)$, and the fourth gives $\nu=5/(3+4x_2-x_4)$. Using (\ref{x342}), both of these results reduce to $\nu={5}/{[2(1+x_2)]}$. Comparing with (\ref{10over3}), these diagrams thus give subdominant contributions. 

\subsection{Scaling Exponent from Absorbing Propagator Scaling into Resummed Vertices}

As a possible scenario for obtaining a more definite prediction from simple assumptions of nonperturbativeness, we add one further conjecture which leads to a different procedure for calculating $\gamma$. This procedure singles out the exponent $\nu=2$ which has been observed in numerical hard-loop simulations \cite{arnoldmoore2,IRS}.

In (\ref{gamma2}) we have used the topological relations (\ref{topoBG}) to write $\gamma_x$
without reference to the number of loops $l$ or internal
propagators $I$. We now assume the  effect of a nonperturbative scaling of resummed propagators to be equivalent to a nontrivial scaling law of vertices such that it corresponds to effectively replacing the
topological constraints (\ref{topoBG}) by
\bea\label{toposubs}
&& l = \frac{1}{2}(2-E+v_3) \to \frac{1}{2}\big[2-E+v_3 + x_3 \nu n_3^\prime\big]\,,\nonumber\\[2mm]
&& I = \frac{1}{2}(-E+3 v_3) \to \frac{1}{2}\big[-E+3 (v_3 +\nu x_3 n_3^\prime)\big]\,,
\eea
and that we can then set $x_2=0$.
We have anticipated the observation (discussed at the end of section \ref{vertexSection}) that higher $n$-point vertices produce subdominant contributions that can be dropped.

This procedure yields:
\bea
\label{gammaEff}
\gamma_{\rm eff} = \nu-4+E\big[1-\frac{1}{2}\nu\big]+\frac{1}{2}\nu v_3+\frac{1}{2}\nu x_3\big[(2+\nu)n_3^\prime -2 n_3\big]\,.
\eea

Having set $x_2=0$, (\ref{x342}) implies $x_3=\frac{1}{2}$, $x_4=1$, which means that resummed vertices are smaller than bare vertices in the infrared. In (\ref{gammaEff}) the terms that contain $n_3'$ and $n_3$ contribute with a positive and a negative sign, respectively, which indicates that effectively large resummed lines compensate the contribution from small resummed vertices. 

For the second diagram in Fig.\ \ref{SD-SE}, using (\ref{zakBareBG}) and (\ref{gammaEff}), with $x_2=0$,  $n_3=n_3^\prime=1$ and $\Delta = -1$ we obtain:
\bea
\label{res}
\gamma_{\rm eff}\big|_{E=2}=1 ~~\Rightarrow ~~ \nu = 2\,,
\eea
in agreement with the result obtained by the numerical simulations in Ref.\
\cite{arnoldmoore2,IRS}.

As a consistency check, we can reconstruct the effective resummed propagator implied by the substitutions (\ref{toposubs}) by writing generally:
\bea
\label{gammaGen}
\gamma_{\rm eff} = (\nu-4) l +\gamma_G I_0 +\tilde\gamma_G \tilde I+\gamma_U (v_3-n_3)+\tilde \gamma_U n_3\,,
\eea
where $I_0$ and $\tilde I$ are the number of bare and resummed lines, respectively, and the parameters $\gamma_G=2$, $\gamma_U=-1$ and $\tilde\gamma_U=-2$ are the scaling exponents for the bare propagator, bare vertex, and resummed vertex, respectively. For the dominant second diagram of Fig.\ \ref{SD-SE}
we have:
\bea
\label{gammaGen2}
1=\gamma_{\rm eff}\big|_{E=2} =\left((\nu-4)\times 1\right)  +(\gamma_{\tilde G}\times 2)+(-1\times1)+(-2\times1) ~~\Rightarrow ~~ \gamma_{\tilde G}=3 \,.
\eea
This is consistent with the Ward identity 
$K\cdot \tilde U(-P-K,K,P) = \tilde G^{-1}(P+K)-\tilde G^{-1}(P)$
which implies $-1 + \tilde \gamma_U=-1-2  = \tilde\gamma_{G^{-1}} = -3$.

\section{Conclusions}

Assuming high occupation numbers such that particle distributions in turbulent cascades can be approximated by their classical limit, we have considered perturbative and nonperturbative scenarios for scaling solutions. Perturbative scaling solutions in nonabelian gauge theories correspond to $2\to2$ processes, with exponent $\nu=5/3$ ($\nu=4/3$) for energy (particle) cascades.

We have also done a non-perturbative analysis and obtained two results using  different senarios. These results must correspond to different, parametrically separated infra-red regimes. This is at least consistent with recent numerical calculations in which different authors have produced both results using methods which may probe different regions of the full momentum phase space.
In the case of maximal nonperturbativeness, namely an effective field theory where the scaling exponent of vertex functions does not become subleading at higher loop orders but has all loop orders contributing equally, we obtained the (model independent) result of $\nu=5$ for energy cascades. This coincides with the nonperturbative infrared fixed point obtained for large-$N$ scalar field theories of Refs.\ \cite{jbLET,Berges:2010ez}, which is however too large to match existing numerical simulations of nonequilibrium nonabelian gauge theories. 

By considering Schwinger-Dyson equations for full vertex functions, where bare and dressed vertices contribute separately with perturbative and nonperturbative scaling exponents, we have found that intermediate values for $\nu$ appear possible. We have proposed a scenario for fixing these scaling exponents in a simple way that yields $\nu=2$, consistent with numerical results from hard-loop simulations of nonabelian plasma instabilities at nonperturbatively large field amplitudes.

\appendix

\section{Scale Invariance of the Flux}
\label{michaSection}

We have calculated the exponent $\nu$ from the requirement of stationary transport, by finding a scaling solution to equation (\ref{Zdefn}). 
In this appendix we show that (\ref{Zdefn}) is equivalent to the requirement that the energy flux be scale invariant, which is the approach used in Ref. \cite{micha}. We calculate the energy flux through a sphere of radius $\Lambda$ and require that the result be independent of $\Lambda$. This flux is given by:
\bea
\label{fluxDefn}
S \sim \int^\Lambda_0 d^3p \,\cdot\,p\,\cdot\,I(p)\,,~~I(p)=\frac{dn(p)}{dt}\,.
\eea
We define the scaling exponent of the collision integral:
\bea
\label{gammaDef}
I(\lambda \, p):=\lambda^{-\theta}I(p)\,.
\eea
To find the momentum dependence of the collision integral we choose $\lambda\sim 1/p$ and write $I(p)=p^{-\theta} I(1)$. 
Integrating we find:
\bea
S(P)\sim P^{\tau}\;\left(\frac{I(1)}{\tau}\right)\,,~~\tau = 4-\theta\,.
\eea
For scale invariant flux we require:
\bea
\label{fluxSoln}
\tau = 0~~~{\rm and}~~~\lim_{\tau \to 0}~\frac{I(1)}{\tau} = {\rm finite}\,,
\eea
which means that the collision integral must have a zero at $\tau$=0.

Using equations (\ref{Zdefn}) and (\ref{fluxDefn}) we have:
\bea
\label{ZI}
Z=\theta(p_0)\int d^3p\,~ \rho(P)\, p\,I(p)\,,
\eea
and therefore it appears that if $I(p,t)$ has a zero at some value of $\nu$ then so does $Z$. However, while this is true for algebraic solutions, like the thermal equilibrium solution $\nu=1$, it does not mean that a scaling solution of $Z=0$ is a zero of the collision integral. 
To see this note that if we  multiply the integrand in  (\ref{2loop}) by a function $f^\prime(P,Q,L)$ that is symmetric under the interchange of any two variables, we would obtain a different critical exponent. 
From (\ref{fluxDefn}) and (\ref{ZI}) we obtain that the scaling solution of $Z=0$ is given by:
\bea
\label{tau}
-1=\Delta = -2+1+\tau\,.
\eea
The -2 is the scaling dimension of the factor $\rho(P)$, and the 1 is the extra factor introduced by the Zakharov transformation (see the discussion under Eq. (\ref{1loopGEN})). From (\ref{tau}) we have that $\Delta = -1$ is equivalent to $\tau$=0, which shows that the conditions of stationary transport and scale invariant flux are equivalent. 

We note that we would also obtain a solution for scale invariant flux if we remove the factor $(p)$ in equation (\ref{fluxDefn}). Physically, this corresponds to the scale invariance of particle flux. 
From equation (\ref{tau}) it appears that $\tau \to \tau-1$ indicates a scaling solution that corresponds to  $\Delta=0$ corresponding to a particle cascade. 
However, as discussed in section \ref{lloopSection}, 
we do not find the $\Delta=0$ scaling solution in a nonperturbative scenario
unless we add the assumption that only particle number conserving processes $n\leftrightarrow n$ processes contribute.

\end{document}